\documentclass[acmsmall,manuscript,screen,nonacm]{acmart}

\usepackage{stfloats}
\usepackage{tcolorbox}
\usepackage{tablefootnote} %

\AtBeginDocument{%
  }

\begin{document}

\title{BounTCHA: A CAPTCHA Utilizing Boundary Identification in Guided Generative AI-extended Videos}

\author{Lehao Lin}
\email{lehaolin@link.cuhk.edu.cn}
\orcid{0000-0002-9379-2232}
\affiliation{%
  \institution{The Chinese University of Hong Kong, Shenzhen}
  \city{Shenzhen}
  \state{Guangdong}
  \country{China}
}

\author{Ke Wang}
\email{kewang1@link.cuhk.edu.cn}
\orcid{0009-0002-0904-6585}
\affiliation{%
  \institution{The Chinese University of Hong Kong, Shenzhen}
  \city{Shenzhen}
  \state{Guangdong}
  \country{China}
}

\author{Maha Abdallah}
\email{maha.abdallah@lip6.fr}
\orcid{0009-0003-8034-4021}
\affiliation{%
  \institution{Sorbonne Université}
  \city{Paris}
  \country{France}
}

\author{Wei Cai}
\authornote{Wei Cai is the corresponding author (weicaics@uw.edu).}
\email{weicaics@uw.edu}
\orcid{0000-0002-4658-0034}
\affiliation{%
  \institution{University of Washington}
  \city{Tacoma}
  \state{WA}
  \country{USA}}

\begin{abstract}
In recent years, the rapid development of artificial intelligence (AI) especially multi-modal Large Language Models (MLLMs), has enabled it to understand text, images, videos, and other multimedia data, allowing AI systems to execute various tasks based on human-provided prompts. However, AI-powered bots have increasingly been able to bypass most existing CAPTCHA systems, posing significant security threats to web applications. This makes the design of new CAPTCHA mechanisms an urgent priority. We observe that humans are highly sensitive to shifts and abrupt changes in videos, while current AI systems still struggle to comprehend and respond to such situations effectively. Based on this observation, we design and implement BounTCHA, a CAPTCHA mechanism that leverages human perception of boundaries in video transitions and disruptions. By utilizing generative AI's capability to extend original videos with prompts, we introduce unexpected twists and changes to create a pipeline for generating guided short videos for CAPTCHA purposes. We develop a prototype and conduct experiments to collect data on humans' time biases in boundary identification. This data serves as a basis for distinguishing between human users and bots. Additionally, we perform a detailed security analysis of BounTCHA, demonstrating its resilience against various types of attacks. We hope that BounTCHA will act as a robust defense, safeguarding millions of web applications in the AI-driven era.
\end{abstract}




\maketitle

\section{Introduction}
CAPTCHA \cite{guerar2021gotta, ousat2024matter}, an acronym for "Completely Automated Public Turing test to tell Computers and Humans Apart," is a type of test used to verify whether an online user is a human or a bot. As such, CAPTCHAs are sometimes referred to as "reverse Turing tests" \cite{takaya2013reverse}. Users must complete a given task and submit the result, and only after being verified as human can they proceed with further actions within an online application. Typically, CAPTCHA pop-up windows that appear when users attempt to log in or make network requests too quickly. Login page intercepts are designed to hinder bots from the start, while rapid network requests may be interpreted as potential bot activity imitating human actions.

The primary purpose of CAPTCHAs is to prevent web crawlers and bots \cite{kubicek2024automating} that simulate human behavior from sending frequent requests, which can make the operational data of web services more authentic, resistant to Sybil attacks \cite{douceur2002sybil}, and place significant strain on a server’s network. A large volume of such requests may be perceived as a denial-of-service (DoS) attack \cite{jhaveri2012attacks}, and attackers with multiple nodes may launch a distributed denial-of-service (DDoS) attack \cite{gasti2013and, karami2016stress}. This disrupts the normal use of web applications for legitimate human users. 
A study on web bots found that, on average, each website used for the experiment receives over 37K requests per month and more than 50\% of them attempting to bruteforce attacks and exploit vulnerabilities \cite{li2021good}.
Additionally, bots can attack social media content by promoting keywords to get them trending on X (previously Twitter), thus creating false trends that can reach a broad audience  \cite{elmas2023analyzing, hays2023simplistic}.
It is also possible to disrupt the content recommendation systems of social media platforms in order to influence public opinion \cite{chen2024seeing}.
The situations above are not only happened in the centralized web, but also and more severe in the decentralized web \cite{zhang2019double, li2009denial}.
While there are various methods to detect \cite{zhang2022beyond, kang2010large, senol2024double, li2023scan} and mitigate \cite{du2024medusa, brustoloni2002protecting, huang2023efficient, amin2020web} such attacks, CAPTCHAs remain one of the most cost-effective and widely used defenses.

In addition to protecting web applications, CAPTCHA can also be used to improve the reliability of crowdsourced labeling tasks where there is no objectively correct answer \cite{alonso2013human}, as well as its advanced technique, within-task CAPTCHAs for Human Intelligence Data-Driven Enquiries (HIDDEN) \cite{alonso2014crowdsourcing} to hold promise as a valuable approach for assessing worker reliability and potentially enhancing label quality.

However, the weakness of CAPTCHA lies in its vulnerability when an attack bot gains the ability to solve its challenges, hence becoming a so-called CAPTCHA solver,  and rendering it ineffective as a defense mechanism. For instance, early text-based CAPTCHAs can now be easily bypassed using simple Optical Character Recognition (OCR) tools, such as Tesseract \cite{smith2007overview}. As a result, designing new CAPTCHAs is an ongoing challenge, involving a continuous and evolving battle against increasingly sophisticated bot intelligence.

With the recent surge in multi-modal Large Language Models (MLLMs) \cite{iong2024openwebagent, yin2023survey}, artificial intelligence (AI) has gained unprecedented capabilities in handling a wide range of complex tasks, including but not limited to understanding text, multimedia, and generating synthesized conclusions. Furthermore, by granting AI the ability to control computers \cite{ma2024coco}, these systems can now take autonomous actions in the digital world.
If an AI-powered agent could operate a computer and control the browser like a human, techniques like browser fingerprinting \cite{eckersley2010unique, nikiforakis2013cookieless} and canvas fingerprinting \cite{laperdrix2019morellian} in preventing bots will face diminishing effectiveness.
Further, once an AI bot which integrated by MLLMs learns how to overcome CAPTCHA challenges, the current CAPTCHA systems could become obsolete overnight.

In recent years, AI-driven video generation models have emerged in an exponential manner, with examples such as Stable Video Diffusion (SVD) \cite{blattmann2023stable}, Sora\footnote{https://openai.com/index/sora/} \cite{liu2024sora}, Pika Labs\footnote{https://pika.art/}, Runway\footnote{https://runwayml.com/} \cite{cui2024feasibility}, Stability AI\footnote{https://stability.ai/}, Kling\footnote{https://kling.kuaishou.com/}, Mora \cite{yuan2024mora} and others \cite{ho2022imagen, ho2022video, luo2023videofusion, khachatryan2023text2video}. 
The videos generated by these models have proliferated on various social media platforms. As model architectures continue to improve and the size of model parameters expands, the length and quality of the generated videos are expected to reach new heights, making it increasingly difficult to distinguish them from real content. In addition to generating videos from prompts and related images, some models are especially capable of video extension (known as video prolongation, and video prediction).
Moreover, these models can customize extended scenes, actions, visuals, and objects within the video using text prompts. They can also create visuals that do not exist in the real world, similar to special effects in films.

\textbf{Motivation.}
In light of the above observations and based on the user study results from the research on unnature changes in video content \cite{henaff2019perceptual} and AI-extended videos \cite{wang2025where}, we argue that the human ability to recognize sudden events, transitions, or anomalies inconsistent with the real world in video content can be harnessed to design a novel web CAPTCHA system. This system would serve as a defense against bot attacks on the internet. The video data used for the CAPTCHA could be generated using AI techniques, specifically by leveraging the extending video capabilities of video generation models.

\textbf{Approach.}
We design a novel CAPTCHA mechanism aimed at defending against current and future, more intelligent web bots and crawlers. This mechanism is based on abrupt changes and transitions in video frames and storylines, where users are asked to identify the points of transition.
Since collecting large datasets of real videos with significant transitions is costly, and the degree of transition in real videos is limited, we use AI video generation models to guide and extend these transitions. As a result, our CAP\underline{TCHA} mechanism relies on human perception to identify guided AI-extended video \underline{boun}daries, and which we name \underline{BounTCHA}. Based on this design, we develop a prototype whereby, after watching the video, users are asked to drag the progress bar to the point where they believe the transition occurs and submit their response. The collected timing data is used to determine the effective range within which humans can recognize the boundary, which then serves as the basis for distinguishing human users from bots. Lastly, we conduct a detailed security analysis to assess the effectiveness of this defense method.

To design a new CAPTCHA, we should follow three basic principles from \cite{chew2004image, gossweiler2009s}: (1) Easy to generate and evaluate. (2) Easy for most people to solve. (3) Difficult for automated bots to solve. 
Therefore, we define several \textbf{Research Questions (RQs)} to investigate the BounTCHA mechanism and help to evaluate it:

\textbf{RQ1:} 
Is it practically feasible to create such a CAPTCHA? (Address the first principle.)

\textbf{RQ2:}
To what extent can humans distinguish the boundary between the original and generated segments of the video? (Address the second principle.)

\textbf{RQ3:}
Is there a potential for this CAPTCHA to be successfully attacked? What is the likelihood of such an attack occurring? (Address the third principle.)

Our \textbf{contributions} can be summarized as follows:
\begin{itemize}
    \item To the best of our knowledge, we are the first to design and implement a novel CAPTCHA based on human perception of AI-extended video boundaries, named BounTCHA. The relevant prompts, working prototype, and videos are open-sourced on GitHub, link: \url{https://github.com/LehaoLin/bountcha}.
    \item We explore the difficulty and cost of creating this CAPTCHA, and examine the range of time biases within which humans can effectively distinguish video boundaries, demonstrating the feasibility of this system as a novel CAPTCHA.
    \item We conduct a security analysis of BounTCHA against various potential attack methods, including random attacks, database attacks, and multi-modal LLM attacks, and demonstrate its effectiveness in defending against robot tasks.
\end{itemize}

The following sections review relevant literature, outline video data preparation, detail the prototype and studies on human performance, discuss the in-depth security analysis, and conclude with limitations and future work.

\section{Related Work}
\subsection{CAPTCHAs}
We have categorized the current mainstream types of CAPTCHA into four groups: text-based, image-based, 3D \& gamified, and video CAPTCHAs.

\begin{figure}[ht]
  \centering
  \includegraphics[width=\linewidth]{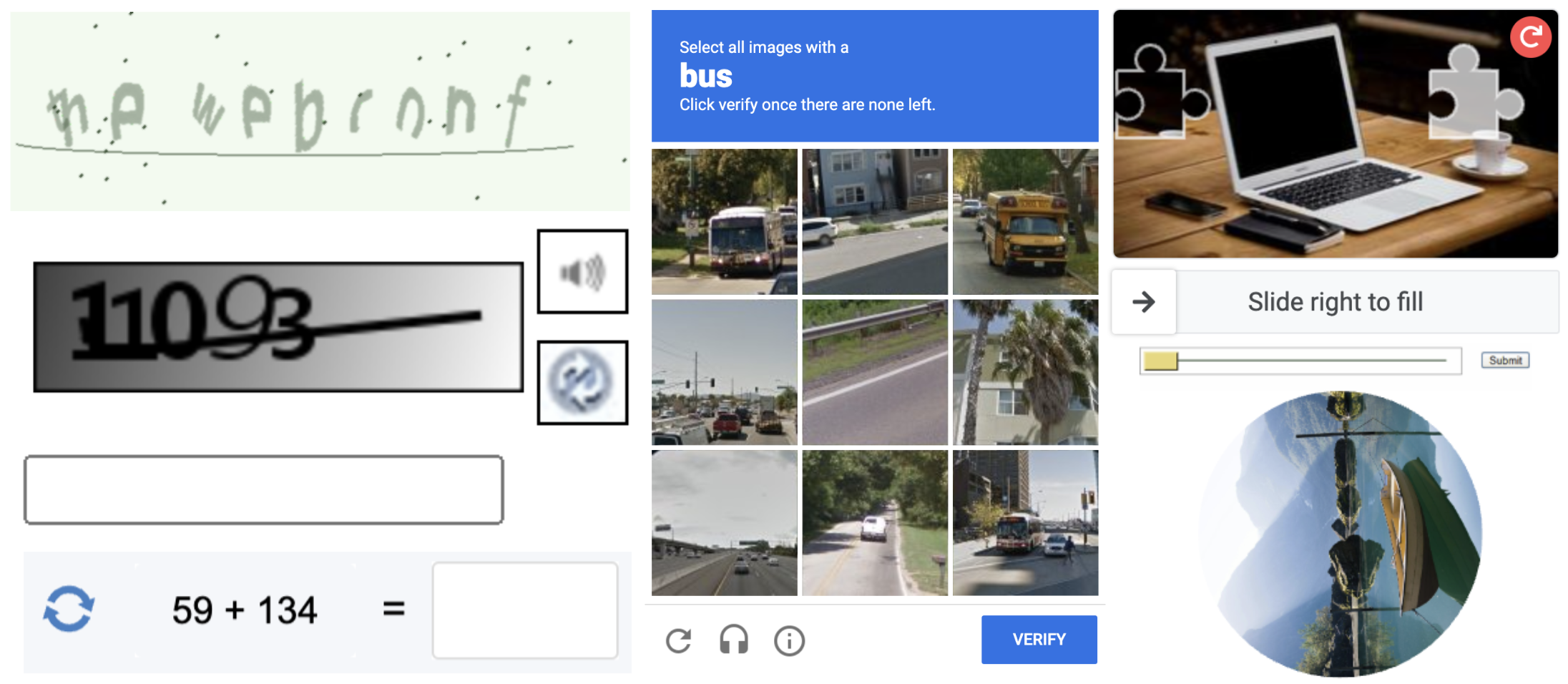}
  \caption{Common text-based and image-based CAPTCHA examples, including arithmetic CAPTCHA, reCAPTCHA, puzzle CAPTCHA, What's Up CATPCHA, and others.}
  \label{captcha-1}
\end{figure}

\subsubsection{\textbf{Text-based CAPTCHAs}} 
Text-based CAPTCHAs have various forms of representation, shown on the left part of Figure \ref{captcha-1}. 
These can be broadly categorized into two types: character-based and digit-based. In the character-based type, distorted text is generated for users to recognize and input the corresponding characters. The digit-based type, in addition to requiring users to recognize and input numbers, may also involve simple math arithmetic operations \cite{hernandez2010pitfalls}. The correctness of the submitted answer is used to verify whether the user is human. Some websites that specifically serve certain countries or regions use their native language as elements \cite{yu2017usability, tende2021development, wang2014multi, ahn2013user}. 
By overlaying text onto background images after applying filters, some provide a word or short phrase and ask the user to click on the corresponding characters in the image in the correct sequence to verify users. 
\cite{yamamoto2010captcha} proposes SS-CAPTCHA, which leverages the human ability to recognise strangeness in translated sentences to detect malware.

\subsubsection{\textbf{Image-based CAPTCHAs}}
Some of these showcases are presented in the middle and right sections of Figure \ref{captcha-1}. Common image-based CAPTCHA mechanisms include multiple image selection, jigsaw puzzles, and image position correction. For instance, the multiple image selection used in reCAPTCHA \cite{von2008recaptcha} can be categorized into selecting the target object from prompts and manually performing semantic segmentation of a large image into a 3x3 grid. A more widely used jigsaw CAPTCHA \cite{gao2010novel, payal2012jigcaptcha} requires the user to slide a piece into its correct position within the image. 
In contrast, the What's Up CAPTCHA \cite{gossweiler2009s} involves adjusting the placement of an image by sliding a bar.
Moreover, the scene tagging \cite{matthews2010scene} tests the ability to recognize a relationship among multiple objects in an image.

\begin{figure}[ht]
  \centering
  \includegraphics[width=\linewidth]{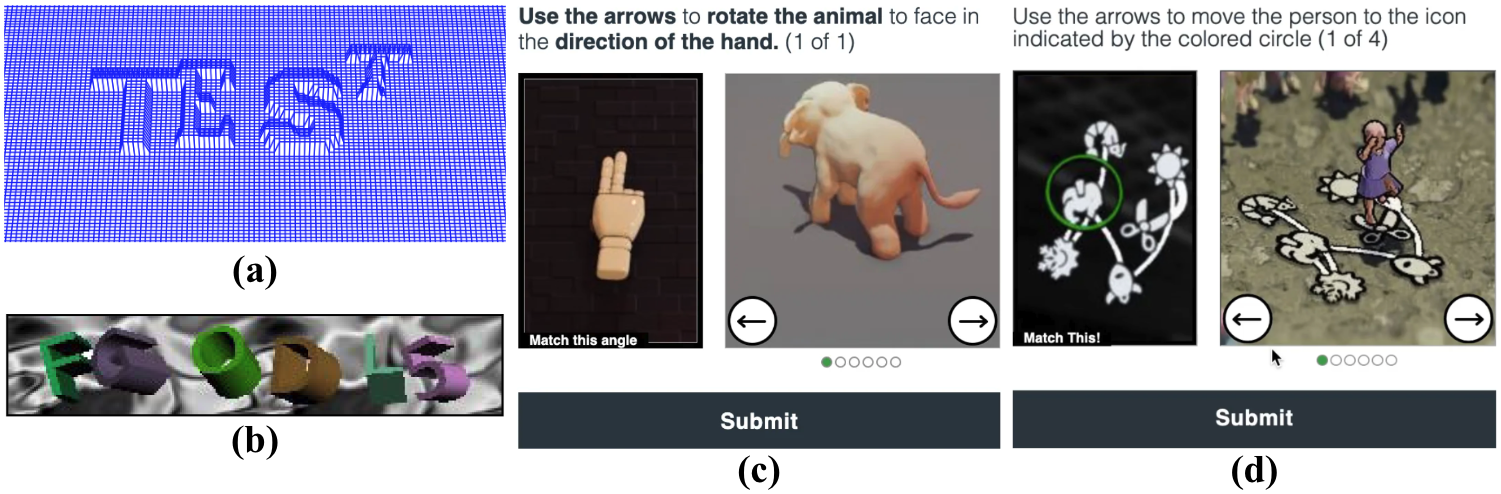}
  \caption{Showcases of 3D \& gamified CAPTCHAs. (a) is a text-based 3D CAPTCHA \cite{nguyen2014security}. (b) is named 3D CAPTCHA \cite{imsamai20103d}. (c) and (d) are gameified CAPTCHAs used by OpenAI's ChatGPT with the 3D view images.}
  \label{captcha-2}
\end{figure}

\subsubsection{\textbf{3D \& Gamified CAPTCHAs}}
Due to considerations of compatibility of 3D object rendering across different browsers and the complexity of generating 3D content, 3D CAPTCHAs often combine features of both text-based \cite{nguyen2014security}
and image-based \cite{ross2010sketcha, woo2017exploration} methods. In Figure \ref{captcha-2}, (a) and (b) illustrate text-based 3D CAPTCHAs, while (c) and (d) demonstrate image-based 3D CAPTCHAs. Additionally, another mechanism, known as gamified CAPTCHAs \cite{mohamed2014three}, incorporates elements of interactivity, as shown in (c) and (d), where users perform various tasks based on the instruction \cite{aggarwal2013animated}. However, with the widespread improvement in hardware and software performance, some CAPTCHA systems opt to disregard browser compatibility and fully leverage modern features. For instance, Dotcha \cite{kim2019dotcha} employs dynamic scatters to represent the 3D effect.

\subsubsection{\textbf{Video-based CAPTCHAs}}
Similarly to our present work, there have been some studies utilizing video as an element in CAPTCHAs. For instance, \cite{kluever2009balancing} uses user-provided descriptions of videos to find relevant labels and tags for verification purposes. \cite{rao2014improved} improves upon this method by replacing text descriptions with a selection-based approach. \cite{anjitha2015captcha} combines text-based CAPTCHAs with video backgrounds, requiring users to input the shown text.

\subsubsection{\textbf{Drawbacks \& Attacks}}
Although the various CAPTCHA forms mentioned above provide varying degrees of protection for web applications, the advancement of AI capabilities has made these defenses increasingly fragile and susceptible to being bypassed \cite{gelernter2016tell}.
For \textbf{text-based CAPTCHAs}, besides traditional Optical Character Recognition (OCR) \cite{chaudhuri2017optical} attacks, even relatively simple machine learning models or neural network architectures \cite{tian2020generic}, such as Support Vector Machine (SVM), K-nearest Neighbors (KNN), and Convolutional Neural Network (CNN), can successfully carry out attacks \cite{chen2017survey, wang2023experimental, starostenko2015breaking}.
For \textbf{image-based CAPTCHAs}, more complex neural network models, such as ResNet \cite{he2016deep} or Vision Transformer (ViT) \cite{dosovitskiy2020image}, can be employed for recognition. Additionally, techniques like edge detection \cite{maini2009study, jing2022recent}, object detection \cite{ren2016faster}, and pixel-level segmentation \cite{long2015fully, chen2018encoder} can be applied to analyze the image, followed by a user interaction simulation programmatically to bypass these defenses \cite{tang2018research, sivakorn2016robot, alqahtani2020image, fritsch2010attacking}.
For \textbf{3D \& gamified CAPTCHAs}, in addition to traditional attack methods \cite{nguyen2012breaking}, since the challenge is still displayed in the form of images on the user interface, attackers can resort to taking screenshots and treating it as an image-based CAPTCHA. They can then leverage multi-modal LLMs to better understand the task and execute the attack \cite{bora2023web, li2015computer, gressel2024discussion}
For \textbf{video-based CAPTCHAs}, there are already numerous vision language models (VLMs) \cite{zhang2024vision, zhou2022learning, bai2023qwen, maaz2023video} capable of interpreting video content and generating textual descriptions. Moreover, this type of CAPTCHA faces challenges related to language internationalization and localization, which limits its usage to specific regions and imposes linguistic barriers, as well as difficulties in understanding complex tasks for users in different education levels.

\subsection{Video Generative Models}

Video generation models possess the ability to modify the generated content in accordance with user input, which is often known as prompts. Based on the kind of prompt, these models can be classified as text-to-video (T2V) models \cite{cho2024sora}, image-to-video (I2V) models \cite{she2022image}, and video-to-video (V2V) models \cite{xing2023survey}. Beyond the straightforward generation of videos, some I2V and V2V models enable fine-tuning, modification, editing, and even the stitching of generated videos by utilizing additional text prompts \cite{hu2022make, sun2024diffusion, chai2023stablevideo, couairon2023videdit}.

In addition to explorations in academic research, video generation technology has been successfully applied in commercial contexts and has reached a relatively mature stage. Companies such as Kling, Runway, Pika , and Stability AI  support text-to-video (T2V) and image-to-video (I2V) generation and provide corresponding API interfaces. According to a report \cite{Upthrust2024} and our practical experience, Kling exhibits superior overall performance in generating videos from images, with more stable output quality. Consequently, we have decided to adopt Kling as the model for our subsequent video generation expansions.

\subsection{Video Understanding}

In the realm of video understanding tasks, video foundation models (VFMs) have embarked on promising endeavors in exploring model architectures, as evidenced by references such as \cite{wang2023all, lei2021less, xu2024pllavaparameterfreellava}. Additionally, they have made significant progress in learning paradigms, \cite{li2023lavender, xu2021videoclip, zellers2021merlot, xiao2024can, yao2024minicpm} serving as examples. As for the latest advancements in video understanding, Tarsier \cite{wang2024tarsier} emerges as a prominent family of large-scale video-language models. These models are meticulously designed to craft high-quality video descriptions. Notably, the model, along with its code and data, has been made publicly accessible for inference, evaluation, and deployment purposes. As of September 3rd, 2024, extensive evaluation results have demonstrated the remarkable capabilities of Tarsier in general video understanding. It has achieved state-of-the-art (SOTA) performance across six open benchmarks, solidifying its position as a leading approach to understand videos. Consequently, in our current research endeavor, we have decided to utilize this model to participate in the prompt generation of AI-extended videos.

\section{Video Data Preparation (RQ1)}
We collected 25 raw videos which are short (3-10 seconds) without significant changes in content from Pexels\footnote{https://pexels.com/, where videos are free to use and allowed to be modified according to their license.}. 
To avoid the unnatural plot twists commonly found in videos (such as anime) and other films as in the science fiction and fantasy genres, which could lead to user misjudgment, we do not select such videos. Instead, we choose those that are captured with real cameras, without any special effects or filters that could cause strange visual distortions.

Moreover, to ensure data diversity while avoiding overwhelming participants in the subsequent user study, which could lead to distorted data, we limit the number of selected videos. Within this constraint, we ensure that the videos feature a variety of shots, scenes, backgrounds, subjects, and other elements to make the collected data as representative as possible.
In this section, we aim to answer RQ1 based on BounTCHA's data preparation.
Next, we describe details of the production pipeline for CAPTCHA video and the video properties before and after processing.

\begin{figure*}[h]
  \centering
  \includegraphics[width=\linewidth]{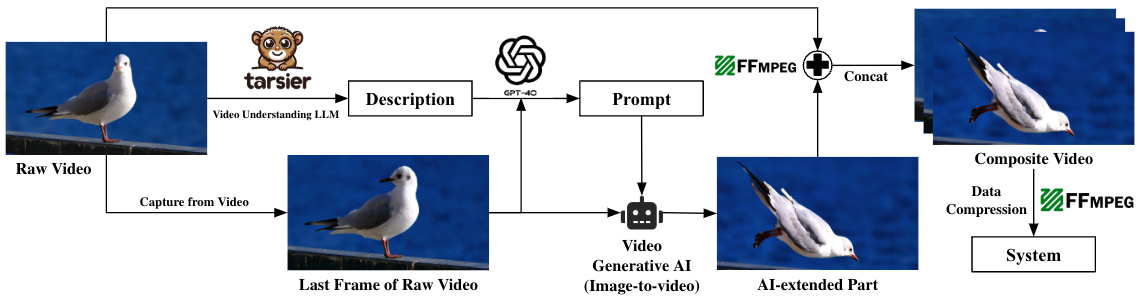}
  \caption{The production pipeline for generating BounTCHA videos.}
  \label{image-to-video}
\end{figure*}

\subsection{Video Generation Pipeline}

The pipeline used for CAPTCHA video production is shown in Figure \ref{image-to-video}. We first use Tarsier to understand the video content and output it as a text description. This description, along with the final frame of the video, is then input into GPT-4o to generate prompts for the video generation AI model. The prompts, along with the final frame of the raw video, are input into the video generation AI to produce the AI-extended part. Next, we merge the raw video with the AI-extended part to create a composite video. Finally, we compress this composite video using FFmpeg to produce the video data used in the CAPTCHA system.

Since Sora is not yet available to the public by September 15th, 2024, and considering the performance and quality of the generated videos (some services do not perform well on video extension tasks), we choose Kling to handle the task of generating extended videos. In addition, given that the data size of the composite video may be too large for CAPTCHA, it could cause significant network strain on the host server in real-world scenarios. Therefore, we use FFmpeg to remove the audio track and compress the video to a target bitrate of 256k, which reduces the network transmission load while ensuring that users can still clearly view the video content.

Therefore, we can denote the raw video as $V_{in} = \{F_1, F_2, ..., F_n\}$, where $F_i$ means $i$th frames and $n$ denotes the number of frames. $LLM_v$ is the video understanding model, and $LLM_t$ is the image-text model. $p$ is the prompt for video understanding. Thus, the prompt for the video generative model is
\begin{equation}
    p' = LLM_{t}(LLM_{v}(V_{in}, p), F_n)
\end{equation}
Then, the extended part is generated as
\begin{equation}
    V_{ext} = \{G_{1}, G_{2}, ..., G_{m}\} = \text{Gen}(F_n, p')
\end{equation}
Finally, the output after concatenation and compression is
\begin{equation}
    V_{out} = \{F^{*}_1,...,F^{*}_n, G^{*}_1,..., G^{*}_m\} = \text{compress}(V_{in} \oplus V_{ext})
\end{equation}
where $F^{*}_n$ can be regarded as the real boundary frame of the output video and is the target for users to identify.

\subsection{Prompts} \label{understanding_prompts}
\subsubsection{\textbf{Prompts for Content Extraction}}
We provide two prompts to assist with video understanding to extract the video's content:
\begin{itemize}
    \item \textbf{Description:} Describe the video in detail, covering all events, actions and camera motions. Also, describe the characters' appearance and the background.
    \item \textbf{Keywords:} Summarize the video with 5 keywords.
\end{itemize}
The video description serves to help the subsequent LLM understand the content of the video, while the five keywords enable the LLM to identify the key elements within the video. This information guides the model in determining which factors should be altered or remain unchanged for the transitions.

\subsubsection{\textbf{Prompts for Video Generation Prompts}}
After obtaining the description and five keywords of the raw video, we incorporate this information into the subsequent video generation prompts, along with the last frame of the raw video. These inputs are then fed into the MLLM to generate a completely new video narrative. This process ensures that the generated prompts significantly differ from the storyline and visuals of the raw video, while still being grounded in its foundational elements. Subsequently, the prompts used for guiding video generation are input into the video generative AI to contribute to the creation of the AI-extended video.

\begin{tcolorbox}[width=\linewidth, colback=white!95!black]

You need to generate a prompt to instruct a video model to generate a subsequent video based on the last frame of a given natural video. You need to ensure that the expanded video differs significantly from the previous video to create a considerable difference between the raw video and the AI-extended part. The focus should be on quickly generating movements that differ from natural laws and common sense, ensuring that humans can react quickly without causing drastic changes in the visuals.

\vspace{1em}

We provide:

1. \textbf{A description of the video:} \{\{ description from Section \ref{understanding_prompts} \}\}

2. \textbf{Five keywords related to the video:} \{\{ keywords from Section \ref{understanding_prompts} \}\}

\vspace{1em}

You need to accept these Requirements:

\vspace{1em}

\textbf{Requirement 1:} The prompt should satisfy: Subject + Background + Movement.  \textbf{Subject}: This refers to the main characters, animals, objects, etc., in the frame. \textbf{Movement}: This indicates the desired trajectory of the target subject. \textbf{Background}: This refers to the setting or environment depicted in the frame. 

 \vspace{1em}

\textbf{Requirement 2:} Generate only a description prompt, within 120 characters, including spaces and punctuation. No additional information is needed. Please answer in English.

\vspace{1em}

\textbf{Requirement 3:} You need to carefully read the "A description of the video" and the keywords. Your expansion should be based on them. Try not to deviate too much.

\end{tcolorbox}

\subsection{Video Quality \& Shift Cutting}

\begin{figure}[ht]
  \centering
  \includegraphics[width=0.99\linewidth]{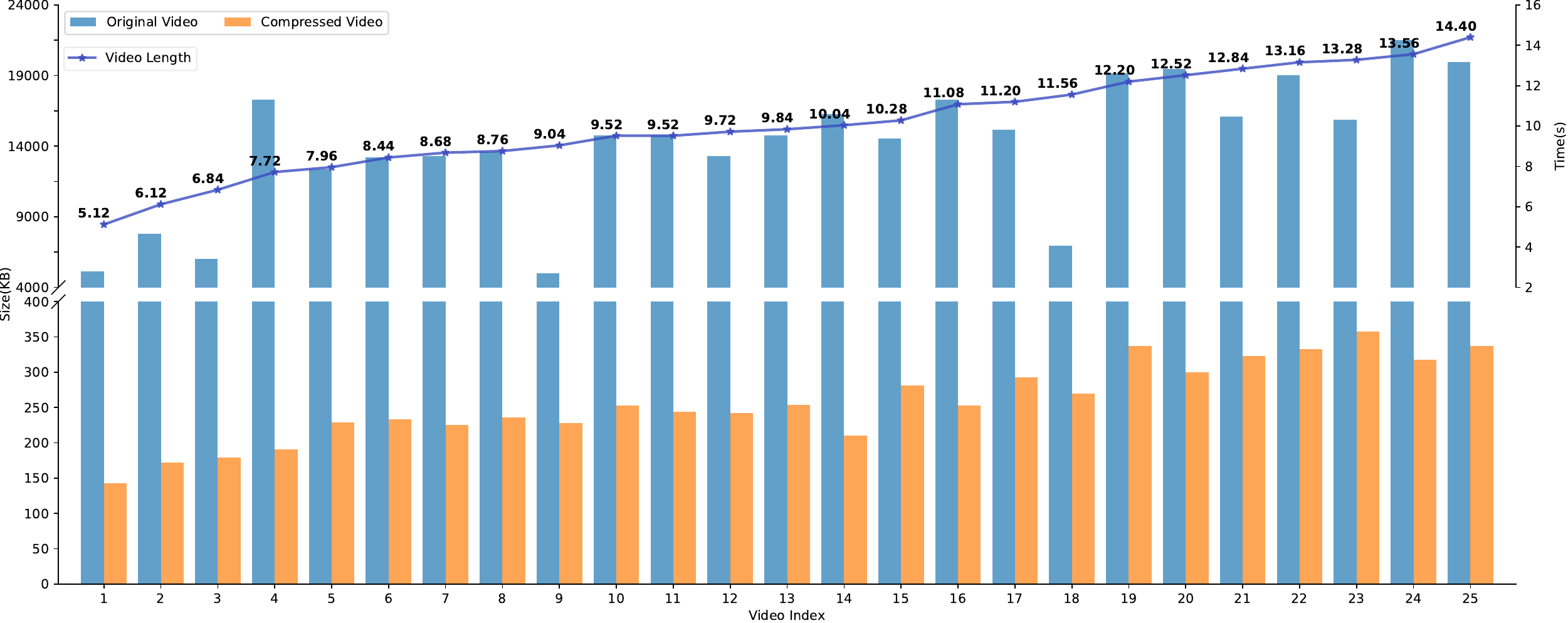}
  \caption{A bar chart comparing the sizes  of original videos ($\mu=14106.62,\sigma=4590.93$) and compressed videos ($\mu=257.68,\sigma=55.75$), alongside a line graph depicting video lengths ($\mu=10.00,\sigma=2.42$), with the videos indexed according to their total duration.}
  \label{video_size}
\end{figure}

To ensure that users can complete the verification process efficiently, our videos range from a minimum length of 5.12 seconds to a maximum of 14.4 seconds. Prior to compression, the smallest video was 5 MB; however, after compression, the largest video did not exceed 350 KB. To simulate a realistic CAPTCHA usage environment, we utilize the compressed videos for subsequent experiments.
The sizes and lengths of the videos before and after compression are illustrated in Figure \ref{video_size}. 
Considering the practical context, to minimize the time users spend on CAPTCHA verification, we selected composite AI-extended videos with durations ranging from 5 to 15 seconds as our experimental data. Due to the inherent compression in the MPEG-4 format, the sizes of the original and compressed videos do not increase linearly with the video length. Some minor discrepancies in size and trends were observed; however, the overall trend in video sizes remains consistent. The average size of the compressed videos used is 257.68 KB, which is an acceptable size that will not place significant strain on the network bandwidth.

In addition to the operations mentioned above, we can also use video cutting to adjust the boundary position throughout the entire video duration. This allows the creation of multiple copies of the same video with different boundary positions. In real-world scenarios, this approach helps expand and enhance datasets, enabling them to handle increased traffic and calls.

\subsection{Time Cost of Video Production}

\begin{figure}[ht]
  \centering
  \includegraphics[width=0.7\linewidth]{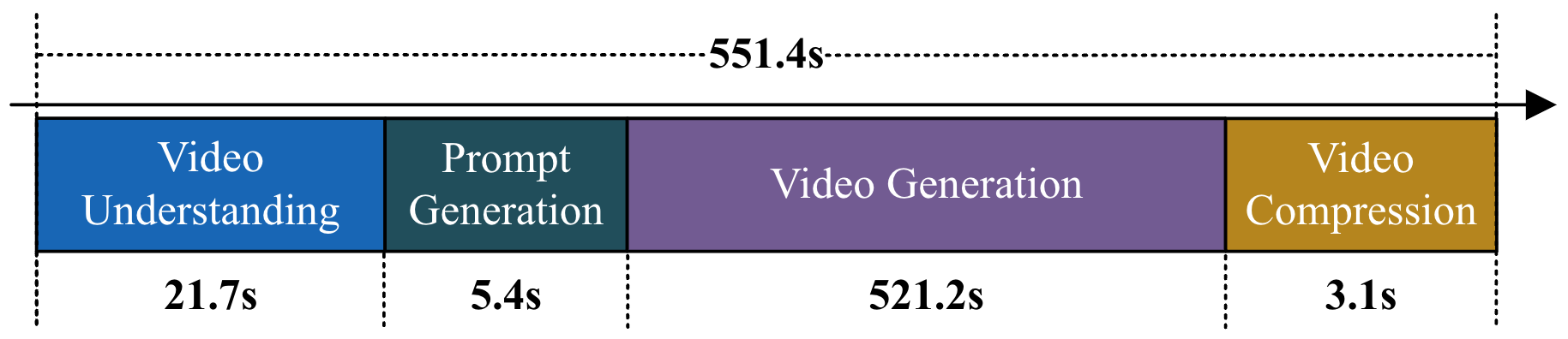}
  \caption{The time cost of the video preparation pipeline. The length of the blocks is not drawn to scale based on the time duration.}
  \label{video_time_cost}
  \vspace{-1em}
\end{figure}

Figure \ref{video_time_cost} illustrates the time consumption in various stages of video preparation, such as video understanding for 21.7 seconds, prompt generation for 5.4 seconds, video generation for 521.2 seconds and video compression for 3.1 seconds.
Although the average time required to generate a video exceeds 500 seconds, multiple variants can be produced rapidly through shift cutting.

\section{BounTCHA Prototype}
\begin{figure}[ht]
  \centering
  \includegraphics[width=0.4\linewidth]{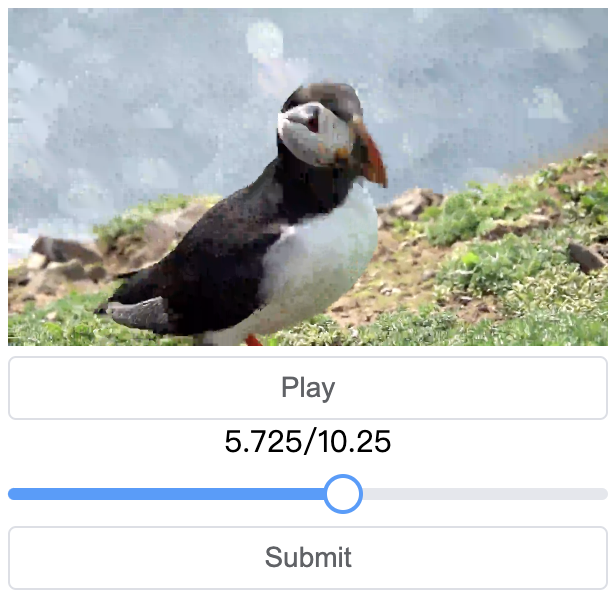}
  \caption{User interface of the BounTCHA prototype.}
  \label{system}
\end{figure}

The system prototype of BounTCHA is shown in Figure \ref{system}. The video playback area is at the top, while the user interaction area is at the bottom. Users can determine the boundary of the composite video by clicking the Play/Pause button and can also drag the slider to seek and adjust the playback. Additionally, users can see the total video length and the current playback time. When users drag the slider to the position they believe marks the boundary, they need to click the Submit button to complete the CAPTCHA. The system then transmits the video ID and the user's confirmed boundary time to the server backend, which records the user's boundary time and compares it with the actual boundary.

The architecture of the prototype is as follows: the frontend is built using Vue.js, the backend utilizes Python with FastAPI, and the database is powered by MongoDB.

In the experimental part of this work, we will gather statistics on the boundaries confirmed by users to determine the effective range of human judgment. In real CAPTCHA scenarios, the server will return the comparison result with the actual boundary. Trials that fall within the effective range are likely to have been completed by a human rather than a robot.

\section{Studies on Human Performance (RQ2)}

\subsection{Experiment}
In this section, we conduct a user study based on the BounTCHA system to explore the discrepancy between users' perceived boundaries of AI-extended videos and the actual boundaries.

\textbf{Setup.}
For ease of both offline and online experiments, we deployed the BounTCHA prototype on a remote server. The server is configured with 2 CPUs and 2GB of memory, running the Ubuntu 22.04.3 LTS operating system.

\textbf{Ethics.}
The school's Institutional Review Board (IRB) reviewed and approved this human-subjects research.

\textbf{Participants.}
We recruited 186 participants for the experiment by posting flyers on campus and advertising on the social media (comprising 83 participants in-person and 103 participants via ZOOM for online experiments). Participants' ages ranged from 18 to 48 years ($\mu$ = 26.93, $\sigma$ = 7.57), and all had prior experience with web-based CAPTCHAs. Each participant compensated with \$0.7 USD.
To ensure data quality and the accuracy of subsequent research, each participant in each experiment is assigned a supervisor. The supervisor will provide training on the experimental procedures before the experiment begins, address any questions the participant may have during the experiment, and monitor the participant’s attentiveness (i.e., ensuring that the answers submitted are not random).

\begin{figure}[ht]
  \centering
  \includegraphics[width=0.6\linewidth]{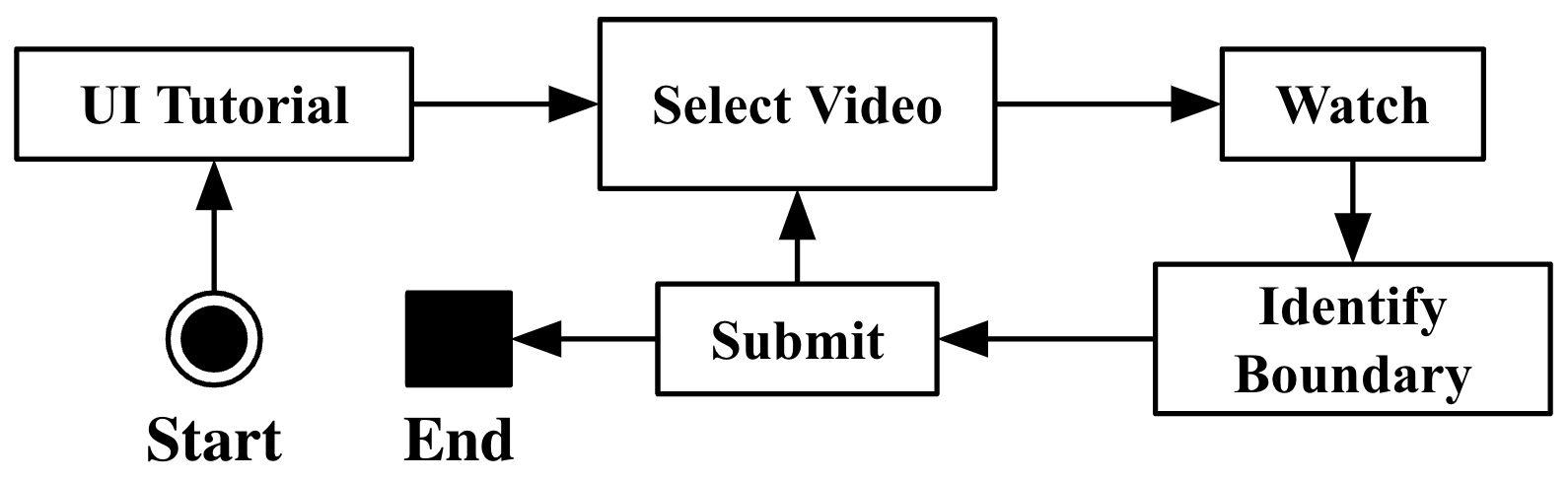}
  \caption{The procedure of user study.}
  \label{procedure}
\end{figure}

\textbf{Procedure.}
As shown in Figure \ref{procedure}, at the beginning of the experiment, we first explained the objectives to the participants and demonstrated how to operate our experimental system.
We informed the participants that the first part is the original video, while the second part is the extended video.
Once the participants became familiar with the system, they sequentially completed a Boundary selection task for 25 videos (in random order). The total completion time for the 25 video experiments was approximately 10 minutes.

\begin{figure*}[ht]
  \centering
  \includegraphics[width=0.9\linewidth]{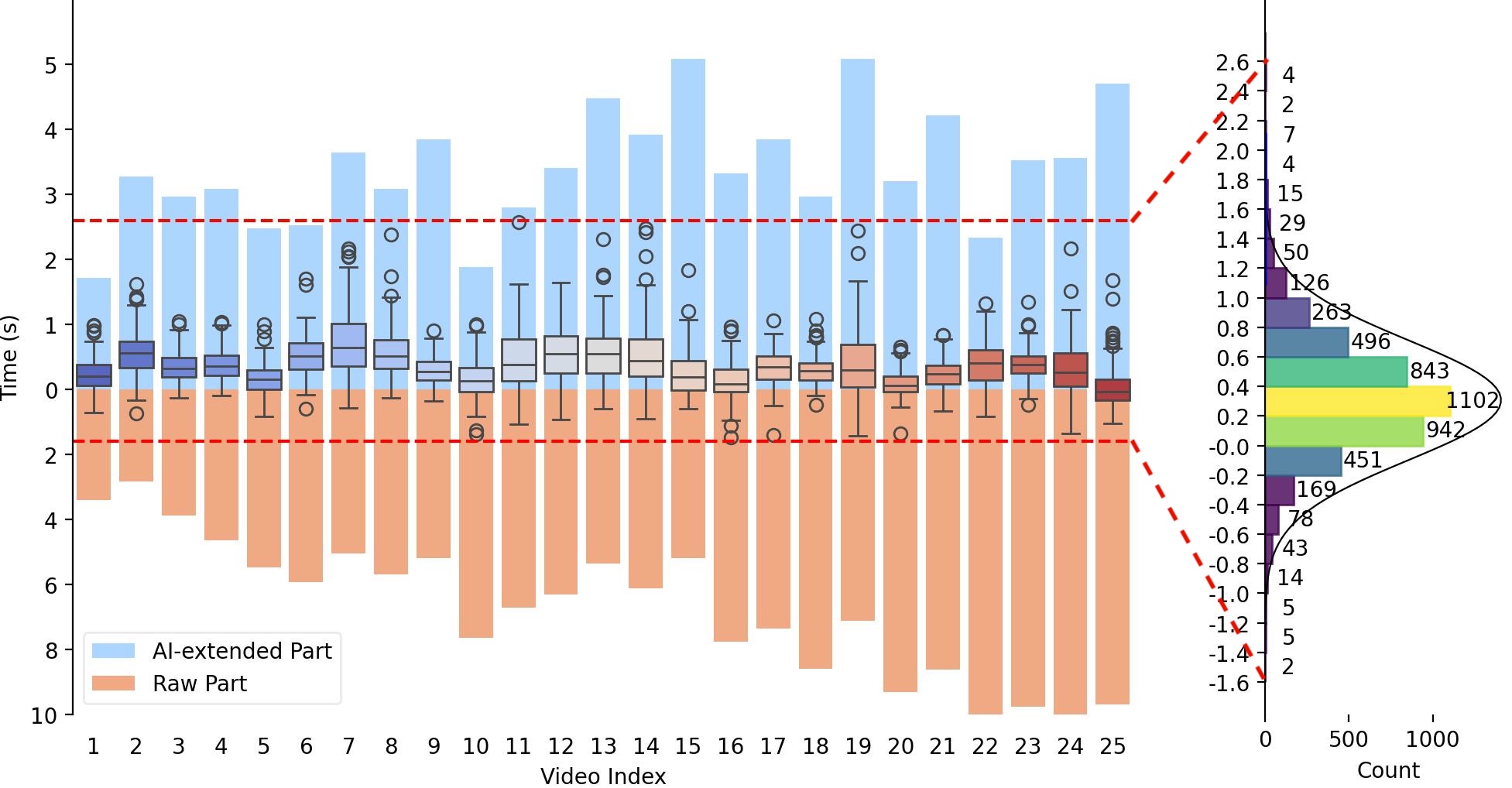}
  \caption{Left: length of videos and distribution of time bias between the human identification boundary and the actual boundary for each video. 0 means the actual boundary between raw video and AI-extended part. Right: the count of all time bias along with its normal distribution estimate ($\mu=0.332$, $\sigma=0.406$).}
  \label{time}
\end{figure*}

\subsection{Results}

\begin{figure}[ht]
  \centering
  \includegraphics[width=0.65\linewidth]{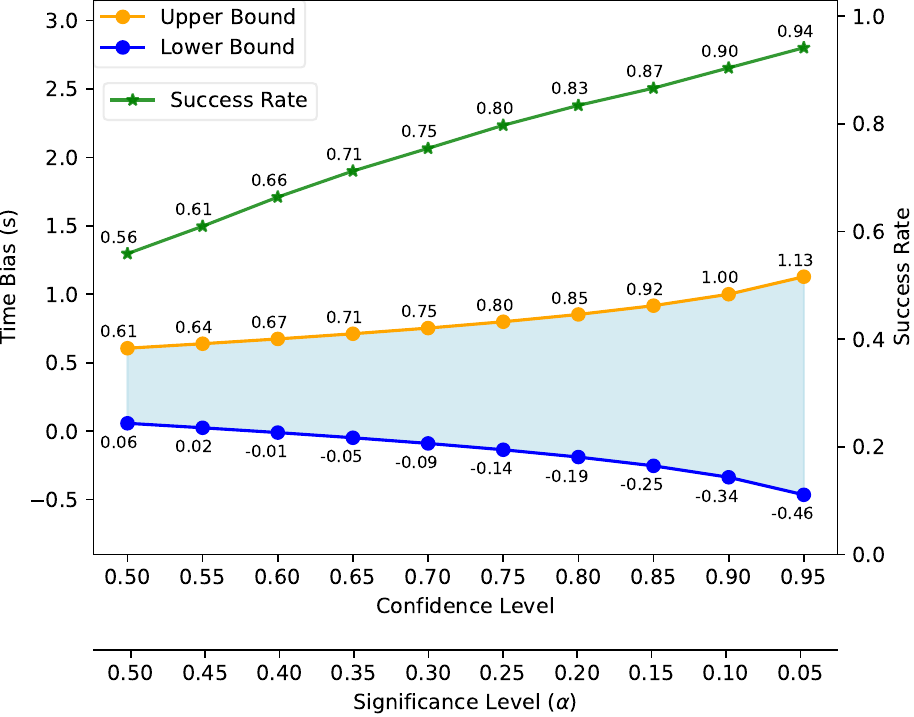}
  \caption{The overall time bias range where the confidence levels with the corresponding significance levels. And their overall success rates from the leave-one-out cross-validation.}
  \label{result2}
\end{figure}

Figure \ref{time} shows the results of the experiment. All the data of time bias falls between -1.6s and 2.6s, and it follows a normal distribution. Therefore, we can increase the difficulty of BounTCHA by adjusting the significance level to narrow the time range.
In Figure \ref{result2}, we show the time bias range where the confidence level varies from 0.5 to 0.95, and the corresponding significance level varies from 0.5 to 0.05. Furthermore, to demonstrate the general applicability of the obtained time bias range, we divide the videos into 5 groups. We derive the range from 4 of the groups and use the remaining 1 group as a validation set to verify the success rate, repeating this for a total of 5 rounds with different validation sets. After the leave-one-out cross-validation, we finally calculate the average success rate, represented by the green curve in Figure \ref{result2}.

\begin{figure}[ht]
  \centering
  \includegraphics[width=0.5\linewidth]{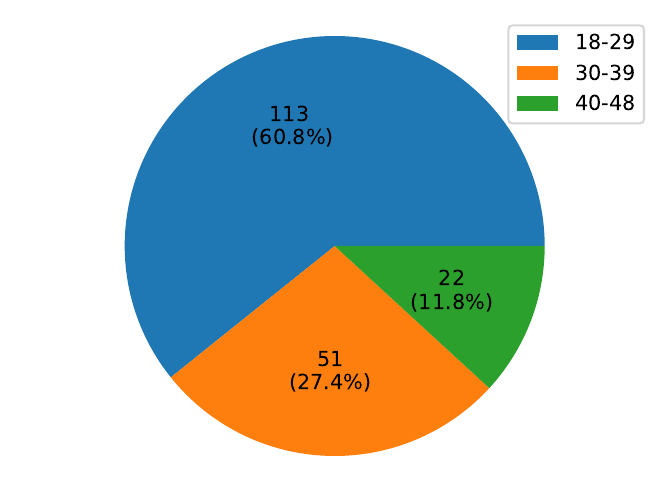}
  \caption{The age group pie chart of the participants.}
  \label{age}
\end{figure}

\begin{figure}[ht]
  \centering
  \includegraphics[width=0.33\linewidth]{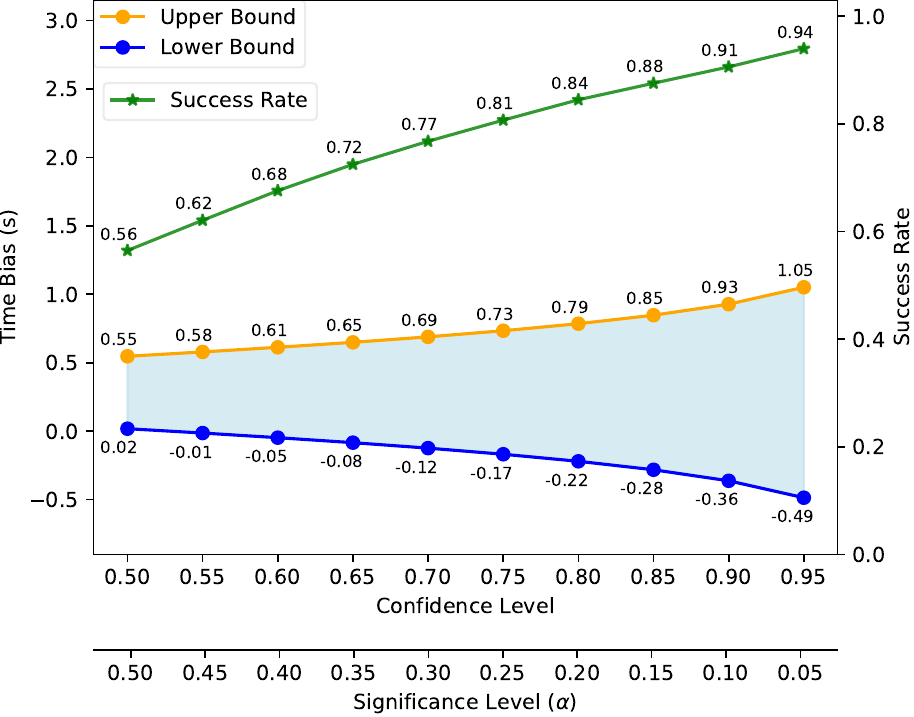}
  \includegraphics[width=0.33\linewidth]{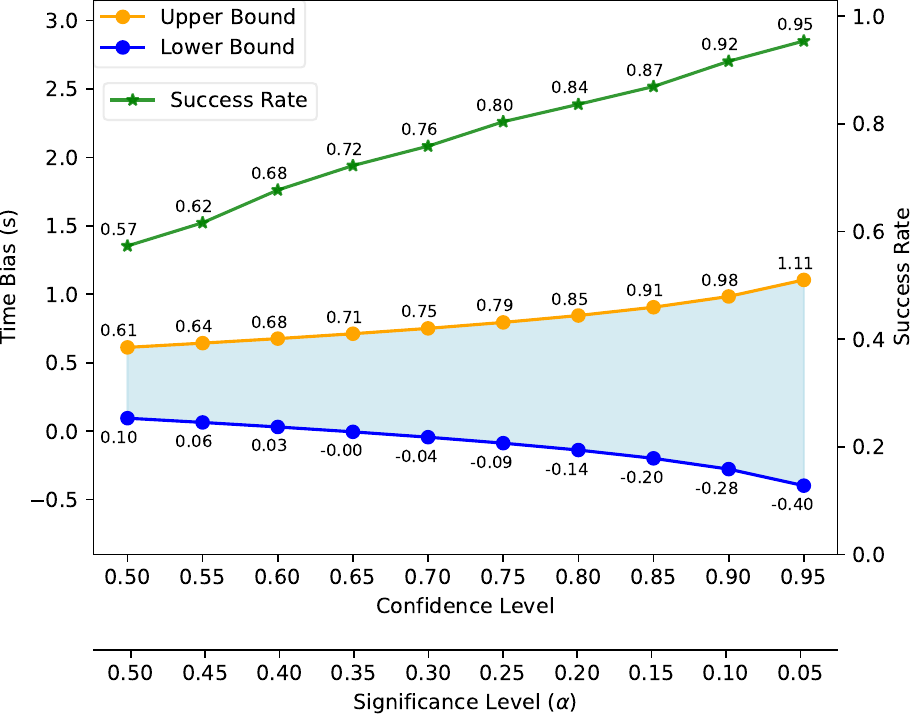}
  \includegraphics[width=0.33\linewidth]{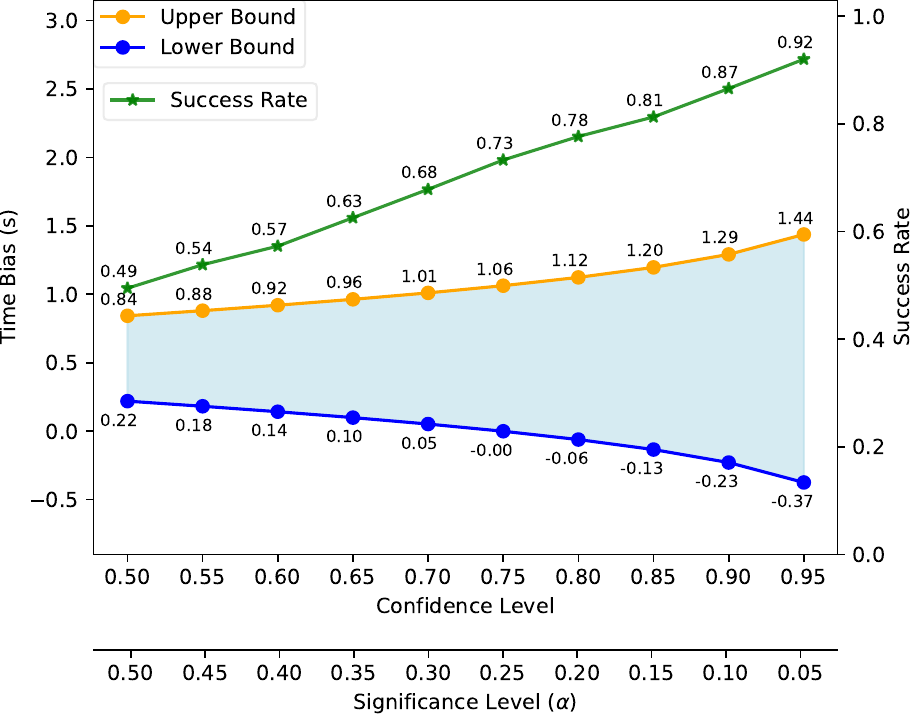}
  \caption{Left: the time bias range and the success rate of the group in the age 18-29. Mid: in the age 30-39. Right: in the age 40-48.}
  \label{result3}
\end{figure}

Based on the age groups presented in Figure \ref{age}, we also analyzed how individuals of different age ranges contributed to the overall participation and success rate. To highlight the differences in success rates among various age groups, we used the overall time bias range as the success criterion, as shown in Figure \ref{result3}. From the results, we observed a subtle pattern: the time bias range tends to be slightly larger for older age groups, while their corresponding success rates are slightly lower. However, the differences between data points at the same confidence level are relatively minor. Therefore, the overall data can be used as the criterion for subsequent analysis. And this section can answer RQ2 from the user study.

Moreover, we use the Pearson correlation coefficient \cite{sedgwick2012pearson} and the Spearman's rank correlation coefficient \cite{zar2005spearman} to assess how the relationship among time bias means \& video lengths, and time bias standard deviations \& video lengths. The calculated results all approach zero, and the p-values are very large. This indicates that there is almost no monotonic relationship between the listed variables.

\begin{table}[h]
\centering
\caption{Correlation Analysis of Bias and Video Length Using Pearson and Spearman Correlation Coefficients}
\label{video_group}
\begin{tabular}{c|c|c}
\toprule
 & \textit{Bias Mean \& Video Length} & \textit{Bias STD \& Video Length}  \\
\midrule
Pearson \cite{sedgwick2012pearson}  & -0.377 ($p=0.062$) & 0.138 ($p=0.510$) \\ \hline
Spearman \cite{zar2005spearman}  & -0.341 ($p=0.095$) & 0.088 ($p=0.675$)  \\ 
\bottomrule
\end{tabular}
\vspace{-1em}
\end{table}

\section{Security Analysis (RQ3)}
We conduct a security analysis based on three attack methods, namely the random attack, the database attack, and the multi-modal LLM attack to answer RQ3. We denote the full duration length of the video as $L$, the attack time bias $x \in X$, the mean value of the time bias as $\mu$, the standard deviation as $\sigma$, the significance level as $\alpha$, the two-tailed confidence level as $1-\alpha$ , the percent point function as $\text{ppf}(\cdot)$ to get z-score, and the effective time bias range $[\beta_{1}, \beta_{2}]$, where
\begin{equation}
    \beta_i = \mu \pm \sigma \cdot \text{ppf}(1-\frac{\alpha}{2}) , i \in \{1, 2\}
\end{equation}

\subsection{Random Attack}

\subsubsection{\textbf{Uniform Distribution Attack}}
\begin{figure}[ht]
  \centering
  \includegraphics[width=0.6\linewidth]{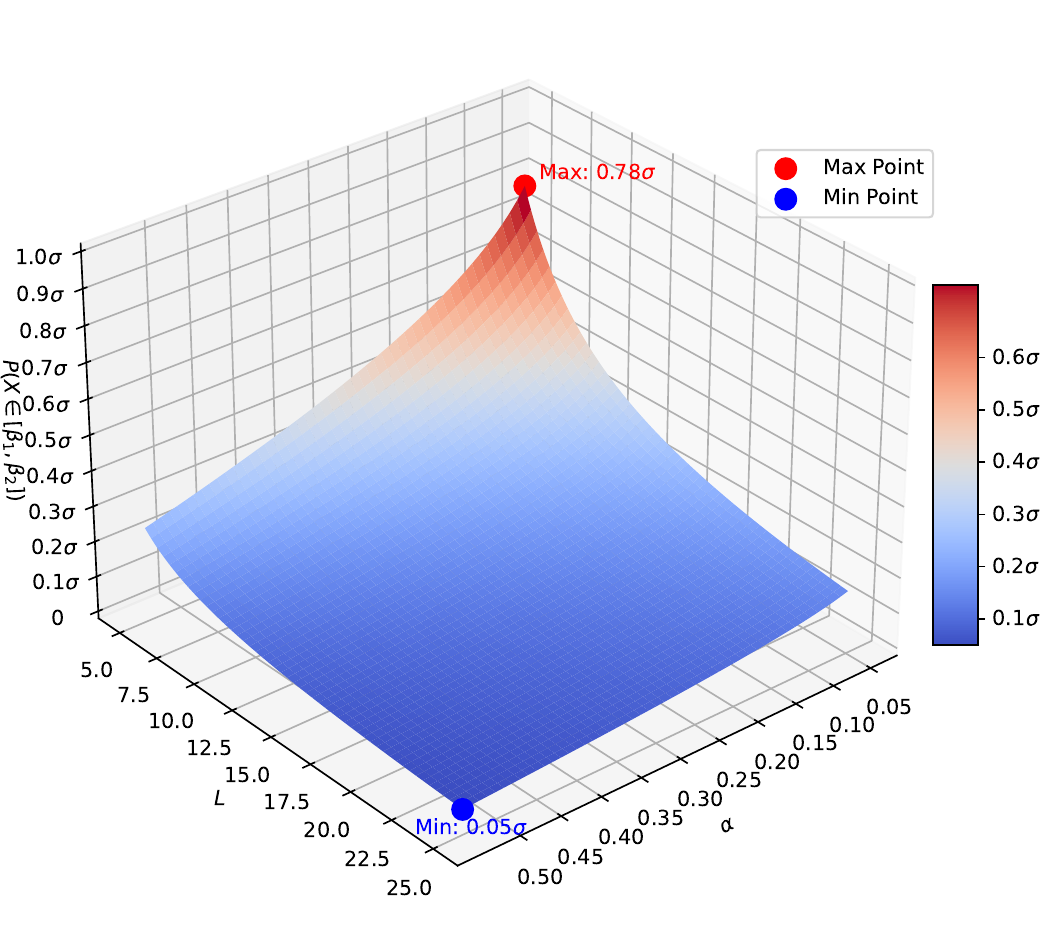}
  \caption{3D Surface Plot of $P(X\in [\beta_1, \beta_2])$: Visualization of the $P(X\in [\beta_1, \beta_2])$ in respect to variables $L$ and $\alpha$.}
  \label{uniform_attack}
\end{figure}

We assume that attackers only know the video length $L$ and use uniform random $X\sim U(0,L)$ to attack BounTCHA. Thus, the attack success probability is
\begin{equation}
    P(X\in [\beta_1, \beta_2]) = \int_{\beta_1}^{\beta_2}\frac{dy}{L} = \frac{\beta_2 - \beta_1}{L} = \frac{2\sigma \cdot \text{ppf}(1-\alpha /2)}{L}
\end{equation}
where $\sigma$ is a constant value computed from the dataset. $L$ and $\alpha$ are variables. Therefore, we can draw a chart to show the relationship among $\alpha$, $L$, and $P(X\in [\beta_1, \beta_2])$, shown as Figure \ref{uniform_attack}.

From the figure, we observe that as $\alpha$ and $L$ increase, the success rate of the attack decreases. Notably, within the range where $\alpha$ is between 0.05 and 0.25 and $L$ is between 5 and 7.5, the success rate declines rapidly. Beyond this range, the rate of decline tends to stabilize. Therefore, to defend against uniform distribution random attacks, it is recommended to set $\alpha \geq 0.25$ and $L \geq 7.5$.

\subsubsection{\textbf{Truncated Normal Distribution Attack}}
We assume that attackers only know the video length $L$ and use the two-tailed truncated normal distribution random to attack BounTCHA, where $\mu' = \frac{L}{2}$, and its probability density function (PDF) is
\begin{equation}
    f(x;\mu',\sigma', 0, L)=\frac{1}{\sigma'}\frac{\varphi(\frac{x-L/2}{\sigma'})}{\Phi(\frac{L/2}{\sigma'}) - \Phi(\frac{-L/2}{\sigma'})}, 0 \leq x \leq L
\end{equation}
\begin{equation}
    \varphi(\xi) = \frac{1}{\sqrt{2\pi}}\text{exp}(-\frac{1}{2}\xi^{2}), \Phi(x) = \frac{1}{2}(1+\text{erf}(x/\sqrt{2}))
\end{equation}
and $f=0$ otherwise, where $\text{erf}(\cdot)$ is the Gauss error function
\begin{equation}
    \text{erf}(z) = \frac{2}{\sqrt{\pi}} \int_{0}^{z}\text{exp}(-t^{2})dt
\end{equation}

Let $\theta_1 = \frac{\beta_1}{L}$ and $\theta_2 = \frac{\beta_2}{L}$ denote proportions of the upper bound and the lower bound with respect to the whole video length, where $0 \leq \theta_1 \leq \theta_2 \leq 1$. So $\theta_2 - \theta_1$ can represent the time bias range length, and $\frac{\theta_1 + \theta_2}{2}$ can represent the middle position in the whole length. The attack success probability is
\begin{equation}
\begin{aligned}
    P(X \in [\beta_1, \beta_2]) &= F(\beta_2;\mu', \sigma', 0, L) - F(\beta_1;\mu', \sigma', 0, L) \\
    &= F(\theta_2; \frac{1}{2}, \sigma'', 0, 1) - F(\theta_1; \frac{1}{2}, \sigma'', 0, 1)
\end{aligned}    
\end{equation}
where $F(\cdot)$ is the cumulative distribution function (CDF).

\begin{figure}[ht]
  \centering
  \includegraphics[width=0.6\linewidth]{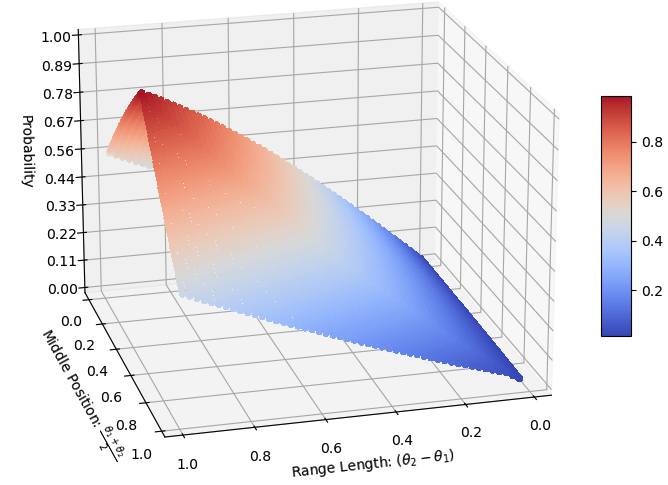}
  \caption{3D Surface Plot of truncated normal distribution random attack success probability in respect to the time bias range proportion and position.}
  \label{normal_attack}
\end{figure}

As shown in Figure \ref{normal_attack}, it is evident that the closer the midpoint of the time bias is to the midpoint of the video duration, the higher the success rate of the truncated normal distribution random attack. Additionally, the shorter the range length proportion, the lower the success rate of bot attacks. From the chart, we can deduce that $\frac{\theta_1 + \theta_2}{2} \leq 0.4$ or $\frac{\theta_1 + \theta_2}{2} \geq 0.6$ with $\theta_2 - \theta_1 \leq 0.5$ is a suitable configuration for the video production.

\subsection{Database Attack}
Inspired by the analysis in the work \cite{ross2010sketcha}, we discuss scenarios where an attacker has partial knowledge of the video database.
For example, the attacker might have limited access to certain videos or some understanding of the general boundary distribution in the video. This consideration allows us to assess the robustness of the proposed BounTCHA system and identify potential vulnerabilities under such informed attack scenarios.

Since we utilize a video cutting method that alters video boundary positions, a single video can have multiple variants. The set of all variants from one single video is referred to as "group". The entire video database is composed of several groups of such variants. Therefore, a BounTCHA video may fall into one of the following three scenarios: (1) the attacker knows the video and its boundary; (2) the attacker does not know this particular video, but is familiar with other variants in the same group and their boundaries; or (3) the attacker is unfamiliar with both the video and any other variants within the same group, as well as their boundaries.

In these three scenarios, the attacker's success probability differs. By comparing frames of the video, the attacker can determine whether the video is fully known. If the video is known, the attacker can leverage existing boundary information to successfully launch an attack. For unknown videos, the attacker can only rely on guesswork. However, when the attacker has knowledge of other variants within the same group, she/he can rule out certain possibilities, thereby increasing the likelihood of a successful guess.

We denote the total number of groups as $M$, with $m$ groups containing at least one variant known by the attacker.
Every video group $i$ has $U_i$ variants, where $i \in \{1,..., M\}$. 
In the $i$-th group, the number of known videos is $u_i$, and $\gamma_i$ refers to the attack success probability.
Additionally, we use $\gamma_{0}$ to represent the success probability that is fully based on guesswork.

The probability $\sum\limits_{i=1}^{M}\gamma_{i}$ of the bot successful attack is
\begin{equation}
    \begin{aligned}
          P  = \frac{\sum\limits_{i=1}^{m}u_i}{\sum\limits_{i=1}^{M}U_i}+\frac{\sum\limits_{i=1}^{m}[(U_i-u_i)\gamma_i]}{\sum\limits_{i=1}^{M}U_i}+\frac{\sum\limits_{i=m+1}^{M}U_i}{\sum\limits_{i=1}^{M}U_i}\gamma_{0}
    \end{aligned}
\end{equation}

Furthermore, we find that $\gamma_i$ is correlated to $u_i$ and $U_i$, which means that if the attacker knows about other variants in the group, the attack success rate is higher. Additionally, we assume that variants' boundaries within one group are uniformly distributed over the video length. Thus, $\gamma_i = \frac{1}{U_i - u_i}$, and the attack success probability is

\begin{equation}
\begin{aligned}
    P=\frac{mu}{MU}+\frac{1}{U}
\end{aligned}    
\end{equation}

Therefore, we formulate the attack probability with $k$ success times in the coming $n$ rounds with the following binomial distribution:

\begin{equation}
\begin{aligned}
    Pr(X=k)=\binom{n}{k}P^{k}(1-P)^{n-k}
\end{aligned}    
\end{equation}

\begin{figure*}[ht]
  \centering
  \includegraphics[width=0.95\linewidth]{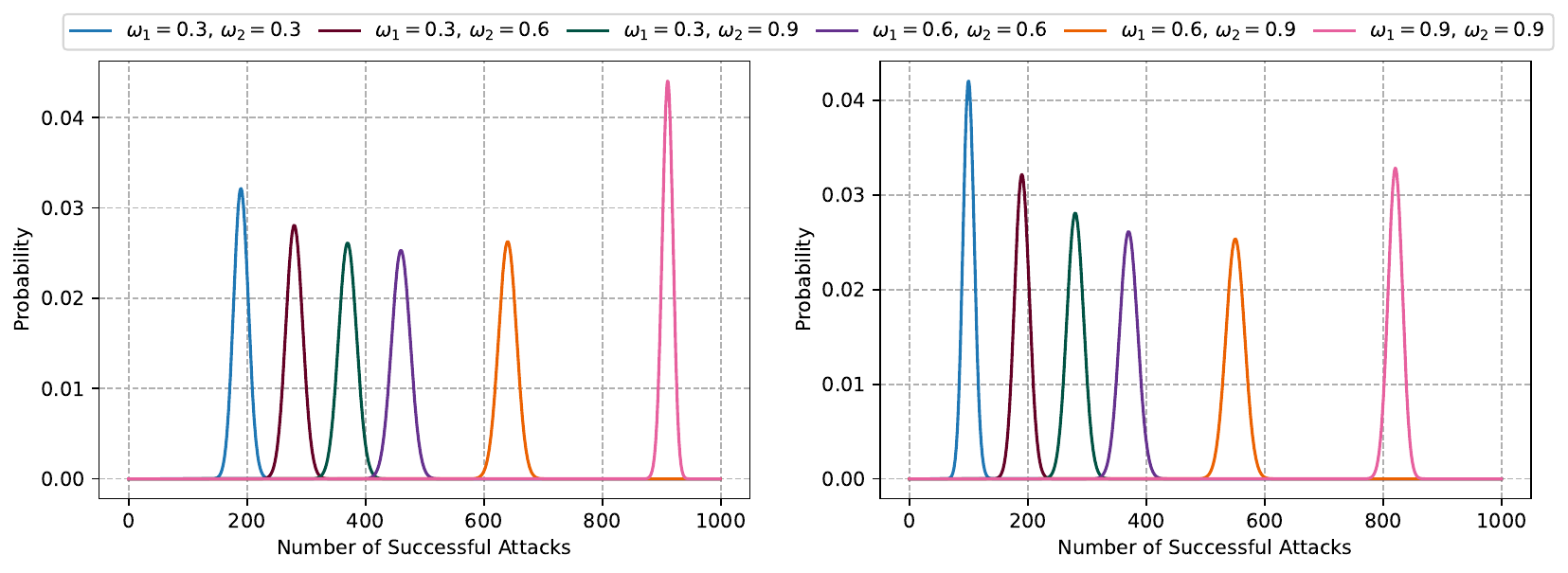}
  \caption{Left: Probability density with $M = 1000$ and $U = 10$, where $\omega_1=\frac{m}{M}$ and $\omega_2=\frac{u}{U}$ ranging from 0.3 to 0.9. Alongside the number of attacks, it shows the corresponding probabilities. Right: Similar to the left, with $M = 1000$ and $U = 100$.
  }
  \label{possion}
\end{figure*}

We assume that the attacker has launched 1,000 attacks. According to the different proportions of data in possession of the attacker to the total data, we have plotted the probability density distribution of the number of successful attacks in Figure \ref{possion}, considering the proportion of the groups containing at least one known variant ($\omega_1$), and the proportion of the total number of video variants known by the attacker across all groups ($\omega_2$). Thus, $P = \omega_1\cdot\omega_2 + \frac{1}{U}$, where $\omega_1$ and $\omega_2$ are symmetrical.
From the examination of each subfigure, our findings indicate that as the values of either $\omega_1$ or $\omega_2$ increase, the number of successful attacks also rises. Additionally, as the curve approaches the extreme values of the x-axis range, the probability distribution becomes narrower and steeper, which means attacks are more stable and have less fluctuation in the situation.
Comparing the differences between the left and right subfigures, we observe that as $U$ increases, the probability distribution shifts leftward. This suggests that as the number of variants within the group grows, the number of successes decreases, which aligns with both common sense and our intuition.

\subsection{Multi-modal LLM (MLLM) Attack}

We tested the recognition capabilities of two open-source MLLMs, Tarsier \cite{wang2024tarsier} and MiniVPM-V 2.6 \cite{yao2024minicpm}, for the boundary of AI-extended Videos. Tarsier is at the state-of-the-art (SOTA) in multiple video question answering benchmarks. MiniVPM-V 2.6 outperforms commercial closed-source models such as GPT-4V, Claude 3.5 Sonnet, and LLaVA-NeXT-Video-34B in Video-MME performance.

\begin{table}[h]
\centering
\caption{The success rate and inference time of the two models for the recognition task.}
\label{llm_attack_rate}
\begin{tabular}{c|c|c|c}
\toprule
 & \textit{Success Rate} & \textit{Time (Mean)} & \textit{Time (Worst)} \\
\midrule
Tarsier-34b  & 13.33\% (10/75) & 20.9s & 38.3s \\ \hline
MiniVPM-V 2.6  &  17.33\% (13/75) &  24.7s & 37.9s \\ \hline
GPT-4V  &   9.33\% (7/75) &  26.4s & 43.2s\\ \hline
Claude 3.5 Sonnet  &  10.67\% (8/75) &  21.2s & 37.1s\\ \hline
Human  & $\geq 82\%$ ($\alpha \leq 0.25$)\tablefootnote{The relevant data is from Figure \ref{result2}.} &  14.2s & 26.6s \\
\bottomrule
\end{tabular}
\end{table}

We conducted three rounds of tests on the two models (Tarsier and MiniVPM-V 2.6) as well as two commercial closed-source models (GPT-4V and Claude 3.5 Sonnet) using our 25 videos. As shown in Table \ref{llm_attack_rate}, in a total of 75 tests, the accuracy rates of both are no more than 20\%, and the running times are all over 20 seconds. The test found that current multimodal large models cannot accurately complete the task of identifying the boundary of AI-extended videos. This is manifested in: inability to understand the task given by the prompt words and returning the frame where the boundary is located instead of video seconds; in some cases, the given video seconds are too different from the real boundary, and in some cases, even the given seconds exceed the total length of the video; there is randomness, as the same prompt words and video will return very different results in multiple rounds of tests. 

\begin{figure}[ht]
  \centering
  \includegraphics[width=0.7\linewidth]{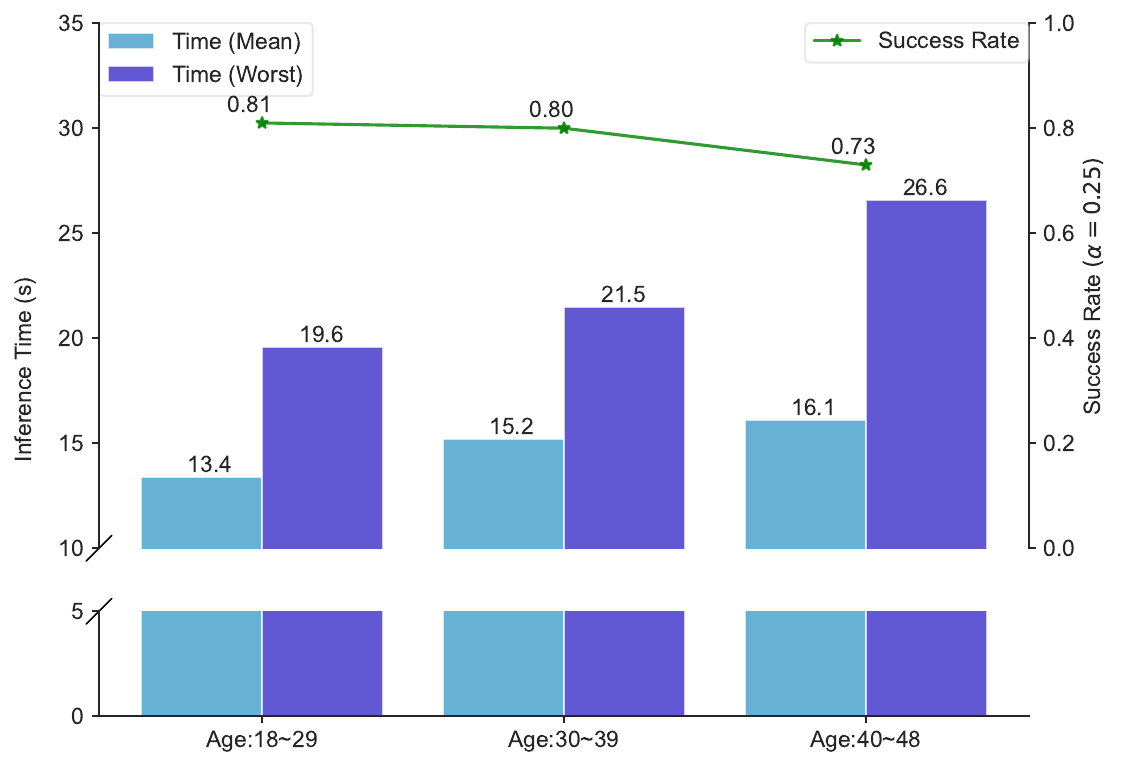}
  \caption{The chart illustrating the relationship between age groups and the average time spent, the longest time spent, and success rates when $\alpha = 0.25$.}
  \label{time_and_srate}
\end{figure}

The time required to complete a verification code is an aspect of human-machine verification. On average, it takes around 10 seconds for a person to solve a typical CAPTCHA \cite{bursztein2010good, xu2020survey}. As shown in Table 2, the average inference time of MLLMs all exceeds 20 seconds. In our experiment, the average time for humans to complete a task is 14.2 seconds, showing a remarkable difference among them.
To provide a better indicator of real-world usability, in addition to the average time, we also recorded worst-case measurements, such as the longest verification time. The human group still exhibited the shortest verification times among all.

Moreover, MLLMs also have high costs. 
Running MLLMs requires a graphics card with more than 16GB of video memory, which has very high requirements for the attacker's equipment.

The chart in Figure \ref{time_and_srate} illustrates the relationship between the average time spent, the time spent on the worst-case scenario, and the success rate across age groups of human participants. We observed that older age groups required more time (both average and worst-case times) and had a lower success rate. The oldest group, in particular, had the longest time spent, which was also notably higher than the worst-case time for other MLLMs. This suggests that human users can currently be distinguished from MLLMs based on time spent when solving BounTCHA. Additionally, the success rates of human users were significantly higher than those of MLLM solvers.

\section{Conclusion}
We designed and implemented a novel CAPTCHA, named BounTCHA, which used guided AI-extended videos as the material, to address the growing threat posed by increasingly intelligent AI-powered bots.
BounTCHA was designed to differentiate between genuine users and bots based on human perception and identification of time boundaries in video transitions and abrupt changes. We developed a prototype of BounTCHA to conduct experiments and determine the effective range of human perceptual time bias, which serves as a basis for distinguishing between real users and bots. A comprehensive security analysis was then performed on BounTCHA, covering random attacks, database attacks, and multi-modal LLM attacks. The results of the analysis demonstrated that BounTCHA effectively defends against various attack vectors. We envision BounTCHA as a robust shield in web security, safeguarding millions of web applications from AI-powered bot threats in an era where machines are becoming increasingly intelligent.

\bibliographystyle{ACM-Reference-Format}

\appendix

\end{document}